\documentclass[a4paper, amsfonts, amssymb, amsmath, reprint, showkeys, nofootinbib, twoside]{revtex4-1}
\usepackage[english]{babel}
\usepackage[utf8]{inputenc}
\usepackage[colorinlistoftodos, color=green!40, prependcaption]{todonotes}
\usepackage{amsthm}
\usepackage{mathtools}
\usepackage{physics}
\usepackage{xcolor}
\usepackage{graphicx}
\usepackage[left=23mm,right=13mm,top=35mm,columnsep=15pt]{geometry} 
\usepackage{adjustbox}
\usepackage{placeins}
\usepackage[T1]{fontenc}
\usepackage{csquotes}
\usepackage{microtype}
\usepackage{booktabs}
\usepackage{tabularx}

\usepackage[pdftex, hidelinks, pdftitle={Accelerating dynamic simulations of photoexcited materials and their evolution by electron-informed machine learning}, pdfauthor={Yunzhe Jia, Fankai Xie, Yunfei Bai, Miao Liu, Cui Zhang, Sheng Meng}]{hyperref}
\begin{document}
\raggedbottom

\title{Accelerating dynamic simulations of photoexcited materials and their evolution by electron-informed machine learning}

\author{Yunzhe Jia}
\affiliation{Beijing National Laboratory for Condensed Matter Physics and Institute of Physics, Chinese Academy of Sciences, Beijing 100190, China}
\affiliation{School of Physical Sciences, University of Chinese Academy of Sciences, Beijing 100049, China}

\author{Fankai Xie}
\affiliation{Beijing National Laboratory for Condensed Matter Physics and Institute of Physics, Chinese Academy of Sciences, Beijing 100190, China}
\affiliation{School of Physical Sciences, University of Chinese Academy of Sciences, Beijing 100049, China}

\author{Yunfei Bai}
\affiliation{Beijing National Laboratory for Condensed Matter Physics and Institute of Physics, Chinese Academy of Sciences, Beijing 100190, China}
\affiliation{School of Physical Sciences, University of Chinese Academy of Sciences, Beijing 100049, China}

\author{Miao Liu}
\email{mliu@iphy.ac.cn}
\affiliation{Beijing National Laboratory for Condensed Matter Physics and Institute of Physics, Chinese Academy of Sciences, Beijing 100190, China}

\author{Cui Zhang}
\email{cuizhang@iphy.ac.cn}
\affiliation{Beijing National Laboratory for Condensed Matter Physics and Institute of Physics, Chinese Academy of Sciences, Beijing 100190, China}
\affiliation{Songshan Lake Materials Laboratory, Dongguan, Guangdong 523808, China}

\author{Sheng Meng}
\email{smeng@iphy.ac.cn}
\affiliation{Beijing National Laboratory for Condensed Matter Physics and Institute of Physics, Chinese Academy of Sciences, Beijing 100190, China}
\affiliation{School of Physical Sciences, University of Chinese Academy of Sciences, Beijing 100049, China}
\affiliation{Songshan Lake Materials Laboratory, Dongguan, Guangdong 523808, China}

\begin{abstract}
Nonadiabatic coupled electron-nuclear dynamics upon electronic excitation underpin the microscopic mechanism and rational modulation of diverse photoinduced functional phenomena in materials, yet their direct first-principles simulations remain computationally demanding. Here we develop a framework for nonadiabatic excited-state machine-learning molecular dynamics (EMLMD) simulations, where the nonequilibrium electronic information upon photoexcitation such as electron temperature is rigorously calibrated from high-precision real-time time-dependent density functional theory (rt-TDDFT) benchmark simulations, enabling accurate reconstruction of excited-state potential energy surfaces (PES). This framework natively incorporates the excited-state electron--phonon couplings and intrinsically captures photoinduced phonon anharmonicity, both of which are missing in standard machine learning molecular dynamics, thus delivering first-principles-level accuracy for excited-state atomic evolutions. Large-scale EMLMD simulations resolve time- and momentum-resolved phonon dynamics in photoexcited materials, directly uncovering the competition between photogenerated coherent phonons and thermal phonons during photoinduced phase transition of bismuth. It also simultaneously resolves elusive atomic-scale microscopic dynamics and global structural rearrangement for selenium photoamorphization. Balancing high accuracy and efficiency, EMLMD offers a versatile paradigm to tackle key challenges in the study of complex excited-state molecular dynamics.
\end{abstract}


\maketitle

\section{Introduction} \label{sec:intro}

Unraveling the dynamical evolution of coupled electron--nuclear systems is fundamental in understanding exotic physical behaviors of materials upon photoexcitation, covering a rich spectrum of light-induced phenomena such as photoinduced superconductivity \cite{Fausti2011, Mitrano2016}, structural phase transitions \cite{Zhang2019, Kundys2010, Xu2022}, transient ferroelectric phases \cite{Nova2019, Li2019}, photo-induced amorphization \cite{Kolobov1995, Vasileiadis2014} and nonthermal melting \cite{Rousse2001}. In state-of-the-art first-principles excited-state simulations, real-time time-dependent density functional theory (rt-TDDFT) combined with surface hopping or Ehrenfest dynamics serves as the primary tool for describing excited-state electronic dynamics and nonadiabatic electron--phonon coupled evolution. In particular, excited-state dynamics is indispensable for characterizing structural evolution with significant atomic displacement, including phase transitions and photochemical reactions, where electron--phonon coupling evolves intricately under electronic excitation \cite{Guan2022, Song2023, Xu2021, You2024, Zhao2025}. Despite its rigorous physical accuracy and substantial advances in revealing the microscopic mechanisms of excited-state electron--phonon dynamics over the past decade, the first-principles framework suffers from prohibitive computational costs. This inherent limitation strictly confines its applications to sub-picosecond time scales and hundred-atom spatial scales, making it incapable of capturing long-time and large-scale physical processes, such as photo-driven long-lived structural relaxation \cite{Klett2018} and the nucleation-and-growth kinetics of phase transition \cite{Sternbach2021}, thereby remaining a critical spatiotemporal bottleneck in this field.

To reduce the high computational cost of first-principles simulations, phenomenological theoretical models have been widely adopted for excited-state dynamic simulations. The two-temperature model (TTM) \cite{Chen2006} is the most popular representative scheme, which simplifies the photoexcited system into two independently thermalized subsystems (electrons and ions) with intrinsic thermal couplings. Although its complexity can be extended by introducing more sophisticated considerations \cite{Carpene2006, Uehlein2022, Jiang2005}, the core assumption of dual thermal equilibrium within a single material remains unchanged. This approximation is feasible for simulating warm dense matter \cite{Norman2013, Zeng2020}, where the electron and ion subsystems exhibit distinct temperatures, while the short-range orders as well as the evolution of atomic-scale local microstructures can be safely neglected. However, it fails to describe crystalline phase transitions accurately, as it ignores crucial spatial correlations and cannot account for the non-thermal electronic distribution prevailing in the early stage of photoexcitation, leading to inherent inaccuracies in modeling light-induced structural evolution of solid materials.

The time-dependent Boltzmann equation (TDBE) approach further optimizes the description of non-thermal electronic behaviors and enables the differentiated characterization of coupling strengths between electrons and individual phonon modes, while simultaneously describing the evolution of both electron and phonon populations. Nevertheless, TDBE still relies on multiple restrictive approximations: (i) it adopts the Markovian approximation based on Fermi's golden rule; (ii) it treats electron--phonon coupling matrix elements as time-invariant constants during excited state evolution; (iii) it employs the harmonic approximation for phonon dynamics, neglecting intrinsic phonon--phonon anharmonic interactions. Numerous experimental and theoretical studies have demonstrated that dynamically evolving electron--phonon couplings and phonon anharmonicity dominate the microscopic evolution of photoexcited materials \cite{Fausti2011, Mitrano2016, Hu2022, Suo2021, Sinner2020}. Consequently, TDBE is unable to reproduce key excited-state dynamic features comprehensively, restricting its application in precise structural dynamic simulations.

Conventional first-principles methods deliver rigorous excited-state nonadiabatic dynamics but cannot scale to large systems over extended timescales, while simplified two-temperature and excited-state decomposition methods fail to capture intrinsic time-variant electron--phonon couplings and full phonon anharmonicity, yielding incomplete microscopic descriptions of nonequilibrium structural evolution. To address this long-standing challenge, here we develop an advanced excited-state machine learning molecular dynamics (EMLMD) framework that embeds time-evolving electronic information (for example, multiple electron temperatures) as an independent dynamic degree of freedom and reconstructs precise excited-state PES from rt-TDDFT benchmark trajectories, as illustrated in Fig.~1. This architecture natively encodes excited-state electron--phonon coupling matrix elements and full phonon--phonon anharmonic interactions, retaining predictive first-principles accuracy while substantially reducing the computational cost of large-scale and long-time excited state dynamic simulations. We systematically apply this approach to two paradigmatic photoinduced transformations in materials across thousands of atoms: light-driven phase transition of bismuth and photoamorphization of selenium. Large-scale EMLMD simulations faithfully reproduce full laser-driven phase transition dynamics of metallic bismuth, resolving time- and momentum-resolved non-$\Gamma$ phonon distributions in photoexcited bismuth and directly uncovering the competitive dynamics between laser-generated coherent phonons and thermal disordered phonons during phase transformation. For the photoamorphization of selenium, our work simultaneously reveals previously inaccessible atomic-scale structural rearrangement pathways and macroscopic global amorphization characteristics.

\begin{figure}[b]
\includegraphics[width=0.96\linewidth]{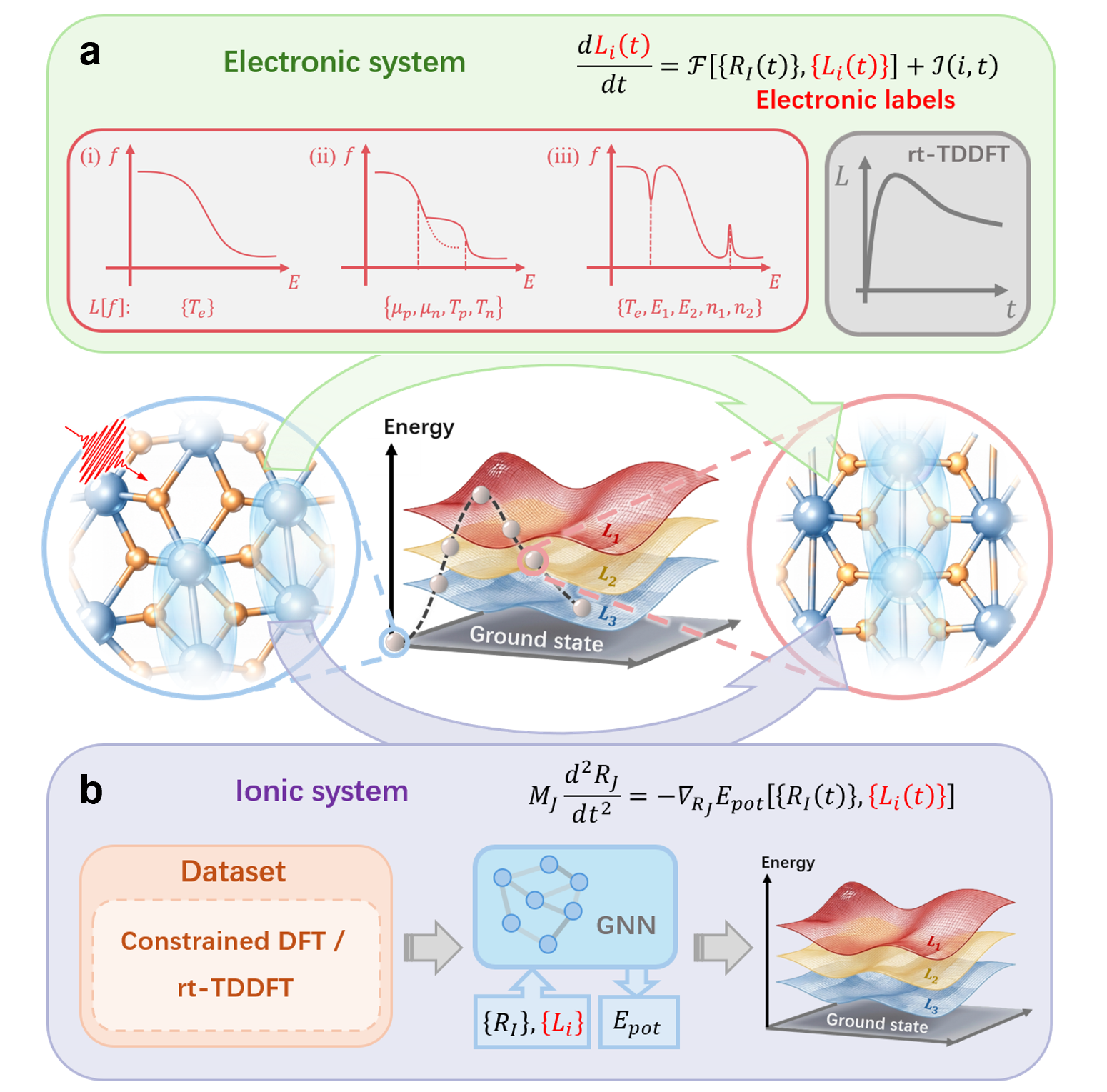}
\caption{\label{fig:1}
\textbf{Schematic diagram of the workflow of excited state machine learning molecular dynamics (EMLMD).} Schematic diagram of an excited-state dynamic process. The illustration shows initial state and an instantaneous state connected by evolution of both electronic and ionic system. \textbf{a--b} The evolution equations and the different scenarios that they might describe for the electronic (\textbf{a}) and ionic (\textbf{b}) subsystems, respectively. $\{\mathbf{R}_I\}$ and $\{L_i\}$ represent atomic coordinates and labels of electronic structure respectively. DFT methods are used to generate training data, while rt-TDDFT trajectories provide the evolution of $\{L_i\}$. Machine learning potential is trained based on self-developed graph neural network (GNN).
}
\end{figure}

\begin{figure*}[!t]
\includegraphics[width=0.8\textwidth]{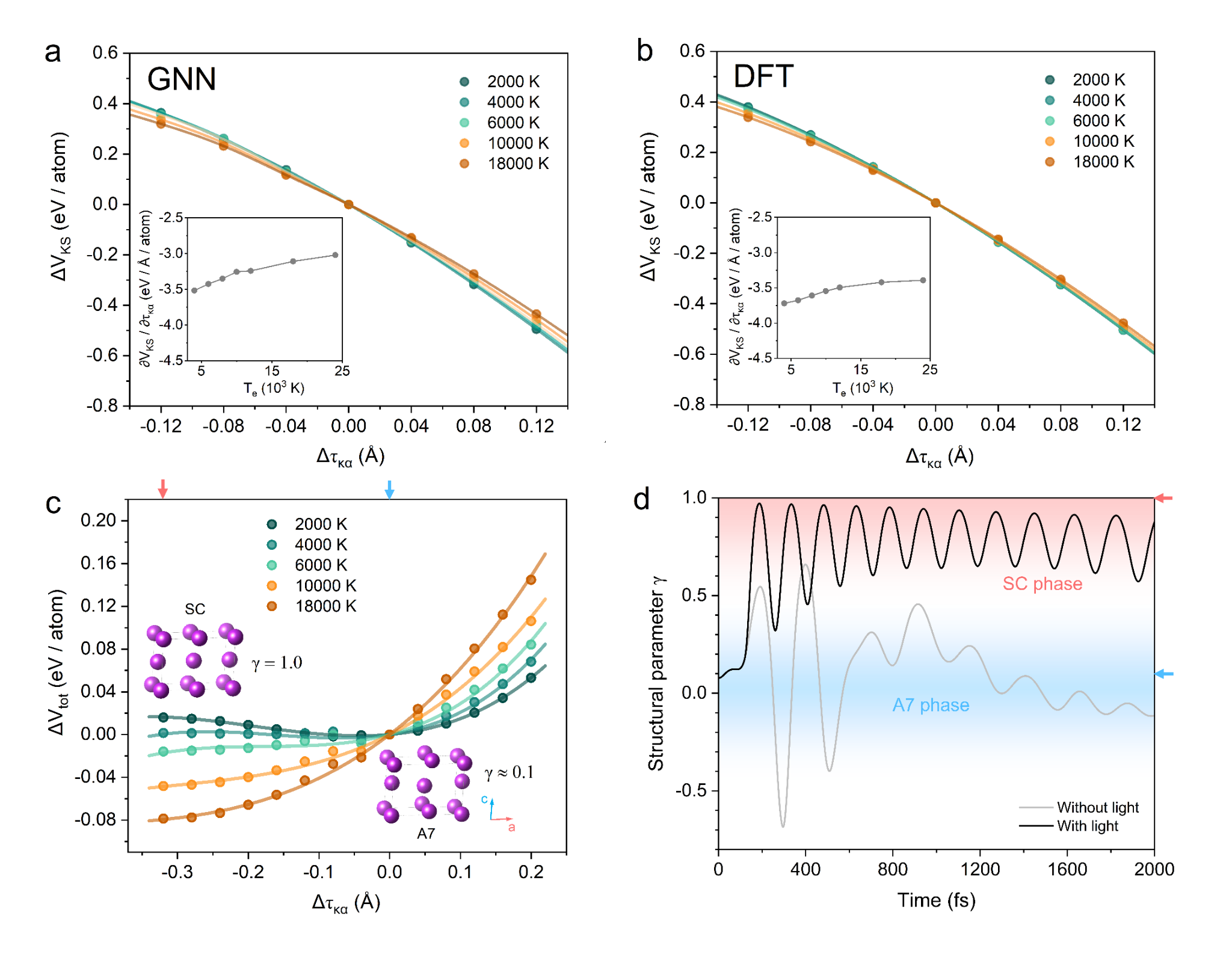}
\caption{\label{fig:2}
\textbf{Machine learning of electron-temperature-dependent excited-state potential energy surfaces.} \textbf{a--b} KS orbital energy variations calculated by the graph neural network (GNN) and DFT, respectively, along the A$_{1g}$ phonon displacement coordinate. Insets show the zero-point ($\Delta\tau_{\kappa\alpha}=0$) derivative evolution as a function of $T_e$. \textbf{c} GNN-predicted total energy profile along the A$_{1g}$ phonon displacement coordinate, with schematic illustrations of A7 and SC lattice structures. \textbf{d} Time evolution of structural order parameter $\gamma$ under unexcited and laser-excited conditions. Black and gray lines represent dynamic trajectories with and without light irradiation, respectively. Red and blue backgrounds denote characteristic SC and A7 phase regimes, respectively, with arrows indicating the ideal structural parameters of each phase.
}
\end{figure*}

\begin{figure*}[!t]
\includegraphics[width=0.9\textwidth]{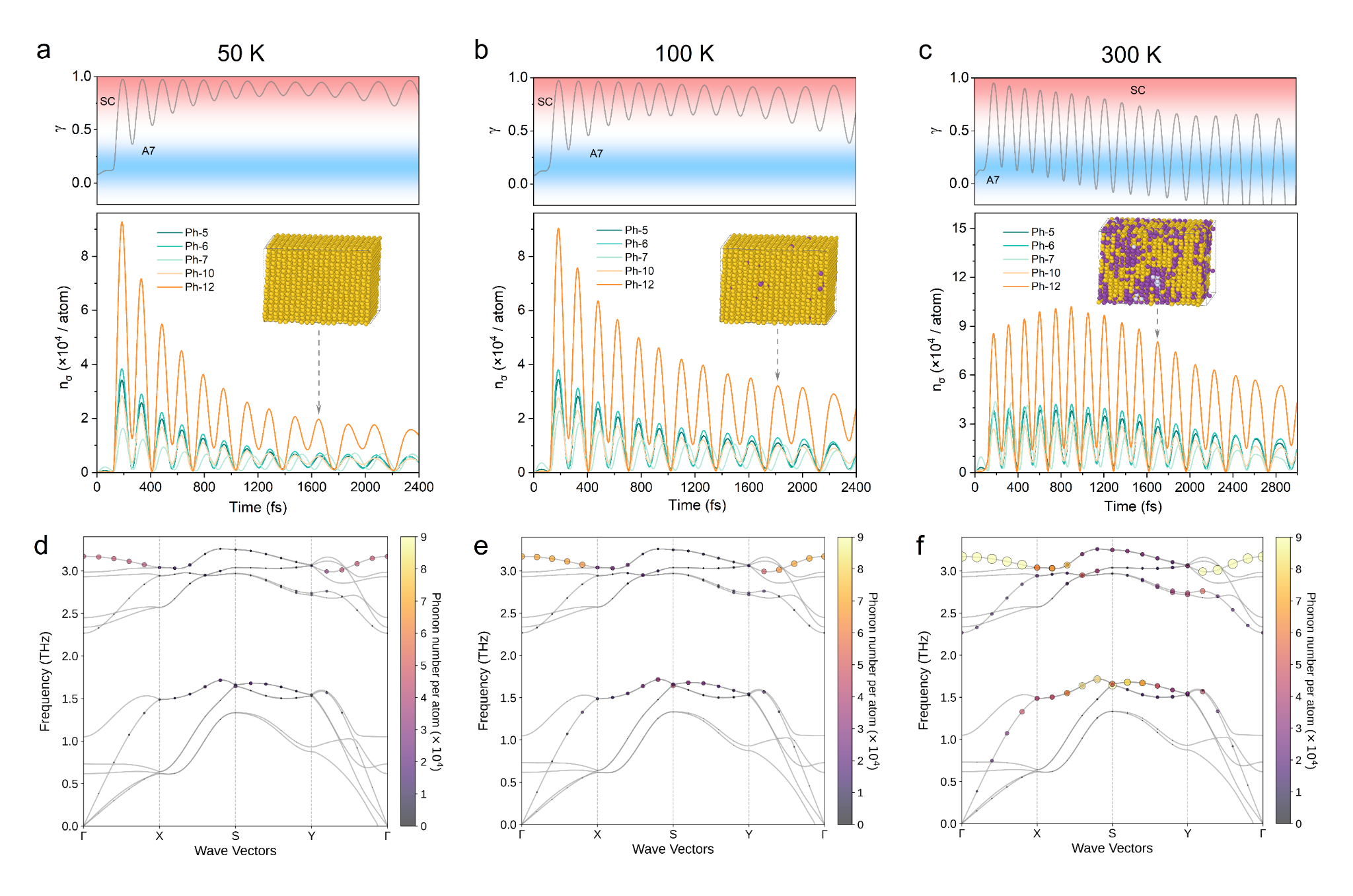}
\caption{\label{fig:3}
\textbf{Environment-temperature-dependent phonon dynamics of photoexcited bismuth.} \textbf{a--c} Time evolution of structural order parameter $\gamma$ (upper panel) and average number of phonons $n_\sigma$ for certain phonon branches (lower panel) at bath temperatures of 50~K, 100~K and 300~K under laser excitation. Red and blue backgrounds delineate SC and A7 phase domains. The A$_{1g}$ mode corresponds to $\Gamma$-point Ph-12. \textbf{d--f} $\mathbf{q}$-resolved phonon occupation distributions per atom at representative time points ($t\sim1700$~fs) near peak phonon density for 50~K, 100~K and 300~K trajectories, with corresponding structural snapshots embedded as insets in (a--c). Circle size and colour encode phonon occupation magnitude.
}
\end{figure*}

\section{Results and Discussion} \label{sec:results}

\subsection{Framework of EMLMD} \label{subsec:framework}

For electron--nuclear systems evolving at the ground and equilibrium states, the atomic structure uniquely determines the free energy of the system, built upon the prevailing Born--Oppenheimer adiabatic approximation. Machine learning molecular dynamics (MLMD) methods \cite{Xie2024, Teitelbaum2018, Giustino2017} can accurately capture ground-state dynamical behaviors by learning the ground-state PES mapped from atomic configurations, achieving computational accuracy comparable to conventional ab initio molecular dynamics (AIMD) when the PES is well-trained.

In contrast, the dynamic evolution of excited and nonequilibrium systems is predominantly governed by electron redistribution, electron--phonon coupling, and phonon--phonon scattering, thereby demanding accurate and efficient dynamical description. The full excited-state PES, which incorporates phase-space information pertaining to electron excitation, can be defined as $E_{\mathrm{pot}}[\{\mathbf{R}_I(t)\},\{\widetilde{L}_i(t)\}]$. Here $\{\mathbf{R}_I(t)\}$ corresponds to time-dependent atomic coordinates, while the finite-dimensional parameter set $\{\widetilde{L}_i(t)\}$ encodes the complete electron distribution information $\{f_{n\mathbf{k}}\}$, where $f$ represents the electron distribution function, while $n$ and $\mathbf{k}$ denote the energy level index and reciprocal lattice vector, respectively. On this rigorously defined excited-state PES, semiclassical ionic dynamics evolve following the Newtonian equation of motion:
\begin{equation}
M_J\frac{d^2\mathbf{R}_J}{dt^2}=-\nabla_{\mathbf{R}_J}E_{\mathrm{pot}}[\{\mathbf{R}_I(t)\},\{\widetilde{L}_i(t)\}].
\end{equation}

A major challenge in constructing excited-state PES lies in the extremely high dimensionality of the electronic phase spaces in condensed matter systems. Explicitly retaining the complete $N_n\cdot N_k$-dimensional electronic information directly leads to computationally intractable dataset collection and neural network training. To resolve this bottleneck, we implement a physically informed dimensionality-reduction scheme that approximates the full electron distribution via a finite-parameter function:
\begin{equation}
f(n,\mathbf{k},t)=\mathcal{L}(\{L_i(t)\},n,\mathbf{k})+\mathcal{O}(n,\mathbf{k},t).
\end{equation}
Here the parameter set $\{L_i(t)\}$ exhibits substantially reduced dimensionality compared with the original high-dimensional electronic phase-space variable $\{\widetilde{L}_i(t)\}$, where $\mathcal{O}(n,\mathbf{k},t)$ denotes the residual fitting error. By adopting physically justified approximations tailored to the system under study, we minimize systematic fitting errors and obtain robust, low-dimensional parameter sets $\{L_i(t)\}$, enabling reliable construction of the excited-state potential functional $E_{\mathrm{pot}}[\{\mathbf{R}_I(t)\},\{L_i(t)\}]$.

For metallic systems, electrons reach approximate thermal equilibrium within $1\text{ ps}$ following photoexcitation \cite{Uehlein2022}, so that the nonequilibrium electron distribution can essentially be parameterized by a single electron temperature, as illustrated in Fig.~1a(i). In contrast, photoexcited semiconductors \cite{Othonos1998} and semimetal materials \cite{Crepaldi2012, Gierz2013} feature distinct Fermi--Dirac distributions for electrons and holes respectively after photoexcitation (Fig.~1a(ii)). Furthermore, systems with structural defects or intense excitation of characteristic energy levels require specific treatment of the occupation states associated with defects and excited energy levels to accurately describe the electron distribution (Fig.~1a(iii)). In Notes 2--3, using VO$_2$ and Se as prototype examples, we present methods for fitting the electron distributions in the second and third types of cases mentioned above. Whether such a level of complexity needs to be considered in actual dynamical simulations depends on the specific scientific questions being addressed.

Full dynamic evolution of excited-state systems requires self-consistent propagation of both ionic configurations (Eq.~1) and electronic states, the evolution of the latter can be expressed by
\begin{equation}
\frac{dL_i(t)}{dt}=\mathcal{F}[\{\mathbf{R}_I(t)\},\{L_i(t)\}]+\mathcal{I}(i,t),
\end{equation}
where $\mathcal{F}$ represents the complex mapping function governing electronic state evolution, and $\mathcal{I}(i,t)$ refers to the residual error derived from the dimensionality reduction approximation, $\mathcal{O}(n,\mathbf{k},t)$. Direct numerical solution of this sophisticated mapping function is non-trivial. Within a region significantly smaller than the beam spot, the material is subject to quasi-uniform spatial irradiation, meaning that the electronic structure evolves in an effectively uniform manner across space. The temporal evolution of electronic parameters is thus calibrated by analysing high-precision rt-TDDFT trajectories obtained from small-scale calculations (see supplementary information (SI) Note~1 for details). The proposed EMLMD framework fundamentally overcomes the spatiotemporal scale limitation inherent to conventional first-principles excited-state dynamic simulations. It enables large-scale atomic dynamical simulations while preserving critical excited-state physical properties, particularly electron--phonon couplings and phonon anharmonic interactions that dominate the microscopic origin of photoinduced dynamic behaviors.

The machine-learning interatomic potential (MLIP) employed in this work is developed based on the Graph-based Pretrained Transformer Force Field (GPTFF) network \cite{Xie2024}. We further refine the network architecture to enhance its capability in capturing subtle structural distortions and fine excited-state structural variations. The training dataset is systematically assembled from extensive high-precision nonadiabatic MD simulations, covering diverse lattice and electron temperatures, equilibrium geometries of stable phase, perturbed transition-state configurations, as well as excited-state trajectories derived from rt-TDDFT calculations. This comprehensive dataset guarantees strong generalization and high accuracy of the resulting interatomic potential for simulating photoinduced dynamics.

\subsection{Photoinduced structural phase transitions} \label{subsec:phase_transition}

To demonstrate the practical applicability of the approach above and the new insights it can offer, we select a few representative materials upon photoexcitation as illustrative examples. Real-time experimental measurements demonstrated the photoinduced phase transition of crystalline bismuth (Bi) from a low-symmetry semimetallic A7 phase to a high-symmetry simple cubic (SC) phase \cite{Teitelbaum2018}. First-principles nonadiabatic simulations based on rt-TDDFT reveal the microscopic nonequilibrium dynamic process underlying such a light-driven Bi structure evolution towards the SC phase (Fig.~S5). Phonon projection analysis reveals that such a structure transition is primarily governed by a high-frequency characteristic phonon eigen mode, i.e., A$_{1g}$ mode (see Fig.~S6). We thus deploy a tailored neural network potential (NNP) based on our validated EMLMD framework (Fig.~S7), which accurately captures the key PES features underlying the photoinduced A7-to-SC phase transition of bismuth.

Notably, our excited-state ML-derived potential inherently incorporates detailed electron--phonon coupling information across excited-state PES, enabling quantitative evaluation of the electron--phonon coupling matrix elements. These elements define the microscopic strength of electron--phonon interaction and are expressed as \cite{Giustino2017}:
\begin{equation}
g_{mn\nu}(\mathbf{k},\mathbf{q})=\left\langle u_{m\mathbf{k}+\mathbf{q}}\middle|\Delta_{\mathbf{q}\nu}v^{\mathrm{KS}}\middle|u_{n\mathbf{k}}\right\rangle.
\end{equation}
Here, $m$ and $n$ denote electronic energy levels, $\nu$ represents individual phonon modes, and $\mathbf{k}$ and $\mathbf{q}$ correspond to electron and phonon wave vectors, respectively. The term $\Delta_{\mathbf{q}\nu}v^{\mathrm{KS}}$ describes the Kohn--Sham (KS) energy variation induced by unit atomic displacement along a specific phonon vibration mode, which can be further decomposed as:
\begin{equation}
\Delta_{\mathbf{q}\nu}v^{\mathrm{KS}}=l_{\mathbf{q}\nu}\sum_{\alpha,s}\sqrt{\frac{M_0}{M_s}}\,e_{s\alpha,\nu}(\mathbf{q})\,\partial_{s\alpha,\mathbf{q}}v^{\mathrm{KS}},
\end{equation}
and
\begin{equation}
\partial_{s\alpha,\mathbf{q}}v^{\mathrm{KS}}=\sum_l e^{-i\mathbf{q}\cdot(\mathbf{r}-\mathbf{R}_l)}\left.\frac{\partial V^{\mathrm{KS}}}{\partial\tau_{s\alpha p}}\right|_{\mathbf{r}-\mathbf{R}_l}.
\end{equation}
In these expressions, $s$ is the atomic index in the primitive cell, $\alpha$ denote the Cartesian directions, and $\tau_{s\alpha p}$ represents the displacement of atom $s$ in the $l$-th supercell along $\alpha$ direction. While $V^{\mathrm{KS}}$ obeys the relation that $\Delta_{\mathbf{q}\nu}V^{\mathrm{KS}}=e^{i\mathbf{q}\cdot\mathbf{r}}\Delta_{\mathbf{q}\nu}v^{\mathrm{KS}}$. Adopting the frozen-phonon approximation, the partial derivative of the KS potential is numerically evaluated via finite atomic displacement:
\begin{equation}
\left.\frac{\partial V^{\mathrm{KS}}}{\partial\tau_{s\alpha}}\right|_{\tau_{sp}^{0}}\cong\frac{V^{\mathrm{KS}}(\mathbf{r};\tau_{s\alpha p}^{0}+\mathbf{b})-V^{\mathrm{KS}}(\mathbf{r};\tau_{s\alpha p}^{0})}{b},
\end{equation}
and $\mathbf{b}\parallel\mathbf{e}_{s,\nu}(\mathbf{q})$. Where $\mathbf{b}$ is the displacement vector applied to atomic coordinates, and $\mathbf{e}_{s,\nu}(\mathbf{q})$ denotes the vibration vector of a certain phonon mode.

To quantitatively characterize the electron-temperature ($T_e$) dependence of electron-phonon couplings, we extracted pure KS potential energy evolution by subtracting nuclear interaction contributions from the total energy (Fig.~2a--b, Fig.~S8). The zero-point slope of each energy curve corresponds to the effective KS potential derivative (insets of Fig.~2a--b), which directly determines the magnitude of electron-phonon coupling matrix element. Notably, our EMLMD scheme faithfully captures the positive correlation between electron-phonon coupling strength and electron temperature, consistent with values from DFT calculations. Beyond electronically modulated electron--phonon interactions, the total potential energy directly determines atomic displacements and intrinsically incorporates phonon anharmonicity. Figure~2c presents the MLIP-predicted total energy $V_{tot}$ along the A$_{1g}$ displacement coordination, which exhibits excellent consistency with benchmark DFT data (Fig.~S8). Increasing $T_e$ gradually shifts the global minimum of PES from the A7-phase toward the SC-phase, demonstrating that photoinduced Bi structural transition is driven predominantly by electronic modulation rather than direct lattice heating.

To elucidate the dynamic evolution of the photoinduced phase transition in bulk bismuth, we perform large-scale MD simulations on a 4000-atom Bi A7-phase supercell (see Fig.~S9) using the well-trained MLIP, under both laser-excited and ground-state conditions. All simulations are conducted at lattice temperatures below 100 K to suppress intrinsic thermal lattice fluctuations. The equilibrium system maintains a constant $T_e$ of 300 K, whereas photoexcited simulations adopted $T_e$ trajectories mimicking high-precision rt-TDDFT calculations (Fig.~S1). Photoexcitation is introduced via a Gaussian-shape laser pulse initiated at 100 fs with a duration of $\sim 64\text{ fs}$ and a peak electric field intensity of $0.386\text{ V/\AA}$.

We define a structural order parameter $\gamma=\langle\gamma_i\rangle$ to quantitatively distinguish the atomic structures between the A7 and SC phases of bismuth, while setting:
\begin{equation}
\gamma_i=1-\alpha\sum_{j\ne k}\min\left\{\cos^4\theta_{jik},\left(\cos\theta_{jik}+1\right)^4\right\}.
\end{equation}
Here $i$ denotes the index of a given atom, and $j$ and $k$ corresponds to its two neighboring atoms. The $\alpha$ is a constant scaling factor used to numerically amplify the differences associated with small structural distortions, and here we set $\alpha=600$. Ideal A7 and SC lattices yield $\gamma\approx0.1$ and $\gamma=1.0$, respectively (Fig.~2c). As shown in Fig.~2d, under equilibrium conditions with constant $T_e$, the Bi lattice undergoes only thermal perturbation variation in $\gamma$, retaining the stable A7 phase. In contrast, rapid elevation of $T_e$ upon laser excitation triggers an ultrafast A7-to-SC structural transition, unambiguously demonstrating that $T_e$ dominates the photoinduced structure transformation pathway.

Notably, large-scale EMLMD simulations further reveal a new competitive interplay between the photoinduced phase transition and lattice thermal perturbation, whereby elevated lattice temperatures destabilize the photo-induced SC phase. Using identical $T_e$ evolution trajectories (Fig.~2d), we perform systematic simulations at bath temperatures of 50 K, 100 K and 300 K. Tracking the temporal evolution of the structure order parameter $\gamma$ in the upper panels of Fig.~3a--c shows that low environment temperatures (50 K and 100 K) enable complete and persistent transition from the A7 phase ($\gamma<0.5$) to the SC phase ($\gamma>0.5$) for over 1 ps. At 300 K, however, intense lattice vibrations induce substantial structural fluctuations, disrupting the metastable SC phase and suppressing photoinduced structural transformation.

To unravel the underlying phonon dynamic origin of the environment-temperature-dependent phase transition behaviors, we implement the phonon projection analysis to quantify photoinduced phonon excitation and coherent evolution (see SI Note~4 for detailed derivation). The amplitude $Q_{\mathbf{q}\sigma}$ of $\sigma$-th phonon mode at wavevector $\mathbf{q}$ follows:
\begin{equation}
Q_{\mathbf{q}\sigma}\propto\sum_{\alpha,s}\left[\sqrt{M_s}\,e_{\mathbf{q}\sigma}^{\alpha}(s)\right]^*\sum_l u_l^{\alpha}(s)e^{-i\mathbf{q}\cdot\mathbf{R}_l}.
\end{equation}
where $M_s$ is relative atomic mass of atom $s$, $u_l^{\alpha}(s)$ denote the displacement of atom $s$ in the $l$-th cell along the $\alpha$ direction, and $e_{\mathbf{q}\sigma}^{\alpha}(s)$ represents the polarization vector of the $\sigma$-th phonon at the $\mathbf{q}$ point, obtained from DFT ab initio calculations. Simulations are performed with a large supercell containing 1000 repeated primitive cells of the A7-phase for large-scale dynamical sampling (Fig.~S9). Within the harmonic approximation, the Hamiltonian of the phonon mode $(\mathbf{q},\sigma)$ is written as
\begin{equation}
H_{\mathbf{q}\sigma}=\frac{1}{2}\left(\dot{Q}_{\mathbf{q}\sigma}^{*}\dot{Q}_{\mathbf{q}\sigma}+\omega_{\sigma}^{2}(\mathbf{q})Q_{\mathbf{q}\sigma}^{*}Q_{\mathbf{q}\sigma}\right).
\end{equation}
The phonon occupation number per-atom $n_{\mathbf{q}\sigma}$ for the $\sigma$-th phonon at the wavevector $\mathbf{q}$ and the average phonon number density $n_\sigma$ for $\sigma$-th phonon branch are defined as:
\begin{equation}
n_{\mathbf{q}\sigma}=\frac{1}{N}\cdot\frac{H_{\mathbf{q}\sigma}}{\hbar\omega_\sigma(\mathbf{q})}=\frac{1}{N}\cdot\frac{\dot{Q}_{\mathbf{q}\sigma}^{*}\dot{Q}_{\mathbf{q}\sigma}+\omega_\sigma^2(\mathbf{q})Q_{\mathbf{q}\sigma}^{*}Q_{\mathbf{q}\sigma}}{2\hbar\omega_\sigma(\mathbf{q})},
\end{equation}
and
\begin{equation}
n_\sigma=\frac{1}{N_q}\sum_{\mathbf{q}}n_{\mathbf{q}\sigma}.
\end{equation}
where $N$ and $N_q$ represent the total number of atoms in supercell and the number of sampled $\mathbf{q}$ points in the primitive cell, respectively.

Time evolution of the individual average phonon number density $n_\sigma$ shown in the lower panels of Fig.~3a--c identifies the A$_{1g}$ mode (Ph-12) as the dominant contributor to photoinduced lattice distortion toward the SC phase. Distinct phonon dynamic responses emerge at different bath temperatures: At 50 K and 100 K, strong electronic excitation substantially activates the A$_{1g}$ mode, driving the system into the metastable SC phase that gradually relaxes back to the A7 phase as $T_e$ decays. At 300 K, persistent thermal lattice fluctuations give rise to sustained oscillations in both phonon occupation and structural order, precluding the formation of SC phase. Structural snapshots captured near the peak phonon number density ($\sim2\text{ ps}$) are presented as insets in Fig.~3a--c. Here, purple color denotes the A7 phase ($-0.5\le\gamma<0.5$), yellow represents the SC phase ($\gamma\ge0.5$), and silver indicates highly distorted lattice structures ($\gamma<-0.5$). The $\mathbf{q}$-point-resolved phonon occupation distributions analysis in Fig.~3d--f reveals highly localized $\Gamma$-point excitation of the key A$_{1g}$ mode (Ph-12), while phonon modes 5 and 6 are preferentially excited near the S-point, demonstrating spatially heterogeneous phonon excitation upon photoirradiation.

To further quantify thermal perturbation effects on phonon coherence, we calculate the phonon coherence factor $\eta$ as derived in SI Note~5. Fully coherent phonons yield $\eta=2$ owing to linear scaling of projection amplitude with supercell size ($Q_{\mathbf{k}}\propto N$), whereas fully randomized thermal phonons converge to $\eta\to1$. Coherence factor analysis in Fig.~S10 reveal that low temperature preserves strong $\Gamma$-point A$_{1g}$ phonon coherence, stabilizing the photoinduced SC phase. At elevated temperature, by contrast, thermal perturbation disrupts the coherent phonon dynamics across most $k$-points, converting coherent photoinduced phonons into incoherent thermal fluctuations and ultimately suppressing formation of the new phase. These findings establish a new competing mechanism: photoinduced coherent phonon excitation drives structural transformation, while thermal lattice perturbation suppresses and disrupts the emergence of light-induced new phase.

\begin{figure*}[!t]
\includegraphics[width=0.9\textwidth]{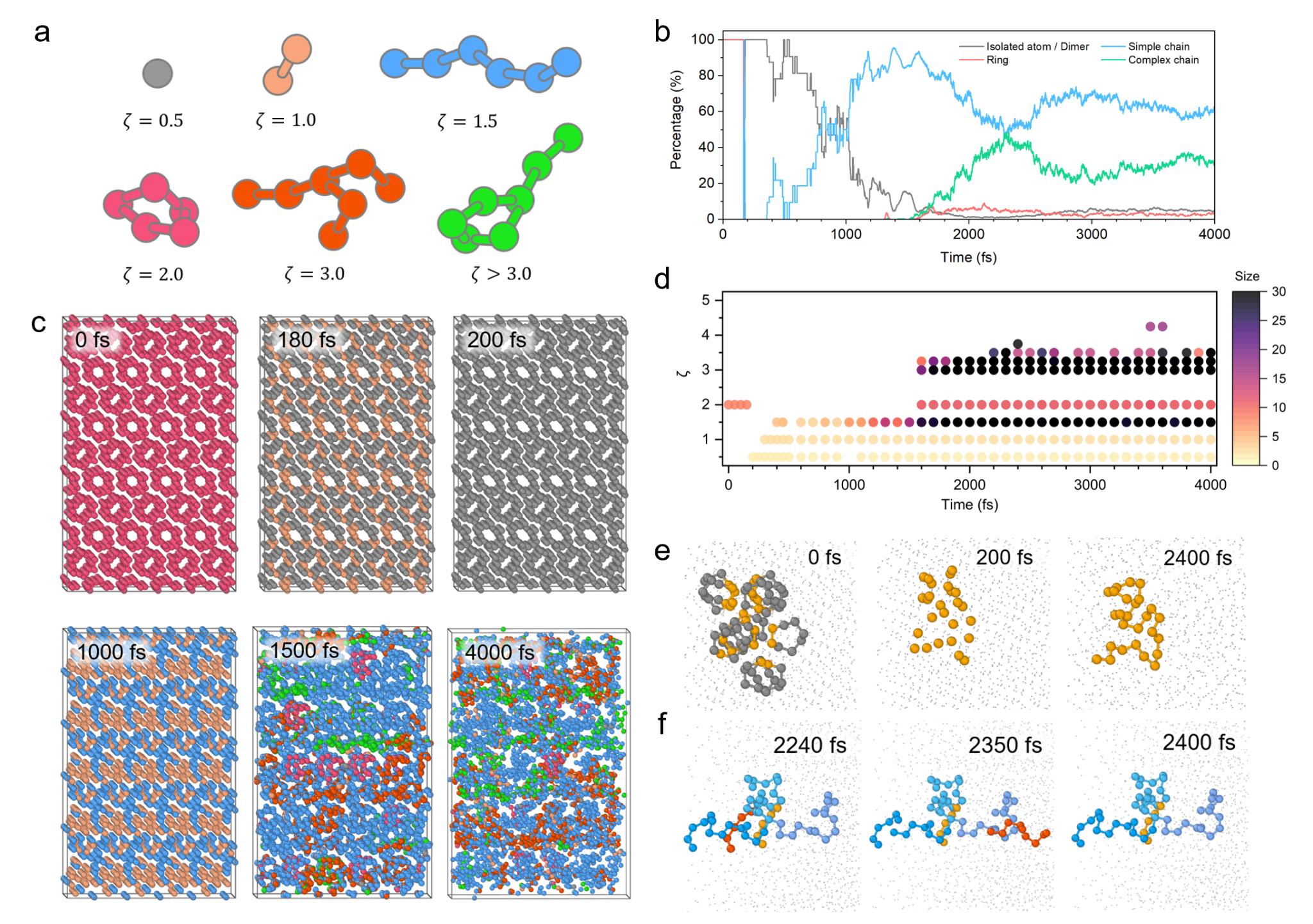}
\caption{\label{fig:4}
\textbf{Photo-induced amorphization dynamics of selenium.}
\textbf{a} Schematic illustration of typical clusters type, with cluster order parameter $\zeta$ values labeled for each structural motif.
\textbf{b} Time evolution of population percentage of distinct cluster species. Grey, blue, red and green lines correspond to isolated atoms/dimers, simple unbranched chains, ring structures and complex branched chains, respectively.
\textbf{c} Representative structural snapshots of system at 0~fs, 180~fs, 200~fs, 1000~fs, 2000~fs and 4000~fs, colored according to the cluster classification criteria in (a).
\textbf{d} Time evolution of cluster types $\zeta$ and sizes throughout the photoinduced amorphization process. Color of each data point denotes cluster size.
\textbf{e} Structure evolution of a representative large cluster identified at 2400~fs. Atoms constituting the 2400~fs cluster are colored orange, while atoms with direct or indirect bonding connections at intermediate time steps are colored gray.
\textbf{f} Structure evolution of a second representative cluster at 2400~fs. Several segments of the main chain are distinguished by adjacent blue hues, while branched sub-chains are colored by adjacent orange hues, respectively.
}
\end{figure*}

\subsection{Photoinduced amorphization} \label{subsec:amorphization}

Glassy chalcogenide semiconductors exhibit prominent structural reconstruction upon photoexcitation, holding a great promise for advanced phase-change memory technologies \cite{Kozyukhin2022, Bogoslovskiy2012, Srivastava2018}. As a prototypical chalcogenide material, the long-standing controversy regarding the structure of amorphous selenium stems from its susceptibility to laser- and heat-induced ring-to-chain structural transitions \cite{Kolobov1998, Roy1998, Lu2024}. In situ extended X-ray absorption fine structure (EXAFS) spectroscopy indicated that photoexcitation increases the average coordination number of selenium through the formation of dynamic interchain bonds \cite{Hegedus2005}. Nevertheless, a comprehensive microscopic understanding of the full dynamic pathway of photoinduced structural amorphization remains elusive. To address this gap, we perform large-scale EMLMD simulations on a 4000-atom homogeneous ring-based monoclinic amorphous selenium (m-$\alpha$Se), systematically elucidating the laser-driven bond rearrangement and underlying mechanism of photoinduced amorphization.

To quantitatively resolve the diverse local structural motifs emerging during dynamic evolution, we develop a coordination-number-based cluster analysis scheme. Adopting a bonding threshold of $2.5\text{ \AA}$ to identify chemically bonded atomic clusters, we define a set of quantitative structural descriptors for systematic cluster configuration classification. We first introduce the parameter $C_n$, which quantifies the number of atoms with a coordination number of $n$ within individual clusters. Further, we define a cluster-type order parameter $\zeta$ to categorize distinct structural units, which is approximately equal to the maximum number of neighbors of an atom in the cluster and reflects the relative structural complexity of the cluster. Simple structural motifs characterized by $C_{n\ge3}=0$ are classified into four categories: isolated atoms ($\zeta=0.5$, $C_2=0$, $C_1=0$), atomic dimers ($\zeta=1.0$, $C_2=0$, $C_1>0$), finite-length chains ($\zeta=1.5$, $C_2>0$, $C_1>0$), and infinite chains or closed rings ($\zeta=2.0$, $C_2>0$, $C_1=0$). A hierarchical classification strategy is further established for complex branched and cyclic architectures. Clusters with minor branching ($C_3>0$, $C_{n\ge4}=0$) yield $\zeta=3.0+0.25\cdot\widetilde{C}_2$ ($\widetilde{C}_2$ is used to merge structures with the same shape but different sizes, see SI Note~6 for detailed definition), while highly branched architectures with four-coordinated nodes ($C_4>0$) correspond to $\zeta=4.0+0.25\cdot C_3$. This quantitative classification scheme achieves precise structural identification and allows for color-coded visualization of all cluster types, as shown in Fig.~4a.

We simulate laser-induced structural evolution of pristine m-$\alpha$Se phase using a Gaussian laser pulse with a peak electric field of $0.257\text{ V/\AA}$, initiated at 100 fs. Statistical analysis of temporal evolution of cluster population reveals a multistage amorphization mechanism (Fig.~4b). Immediately after $T_e$ reaches its maximum, the system undergoes rapid structural breakdown, where the initial cyclic architectures are completely fragmented into isolated atoms or atomic dimers. These fragmented units rapidly recombine into short linear chains within tens of femtoseconds. A mixed structural regime comprising isolated atoms, dimers and simple chains persists from 400 fs to 1200 fs. Beyond 1.2 ps, the population of fragmented units declines drastically, while complex branched chains continuously grow and eventually dominate the system composition. Representative structural snapshots at key dynamic stages, colored according to our standardized cluster classification rules, validate this sequential structural transformation pathway (Fig.~4c).

Dynamic evolution of cluster microstructures and size throughout the amorphization process (Fig.~4d) clearly reveals that primary bond cleavage in the rings initiates at 150--200 fs, followed by spontaneous re-bonding events after 250 fs. No complex branched clusters are observed during 300--1200 fs, and the system is overwhelmingly populated by small clusters containing no more than ten atoms. Large-branched structures emerge after 1200 fs, along with the formation of clusters with more than 25 atoms. To elucidate the atomistic mechanism governing complex structure formation, we track the evolutionary trajectory of individual large clusters. A representative cluster demonstrates that multiple initial ring units undergo sequential bond scission and structural rearrangement to assemble into extended linear chain configurations (Fig.~4e), in line with the experimentally measured decrease in the average coordination number of Se upon photoexcitation \cite{Roy1998}. Complementary tracking of a second cluster reveals frequent reversible connection and disconnection between side branches and main backbones (Fig.~4f), providing direct microscopic pictures for the structural dynamics proposed by Heged\"us et al. \cite{Kolobov1998, Hegedus2005}. On one hand, molecular dynamic simulations can provide evolutionary details that are inaccessible by EXAFS spectroscopy. On the other hand, by specifying an ideal m-$\alpha$-phase initial state, we can capture a much clearer, atomistic scale view of ring-to-chain transformation process during photoamorphization, whereas experiments inevitably deal with a far more intricate amorphous-to-amorphous evolution.

\begin{table*}[!t]
\centering
\caption{\label{tab:method_comparison}\textbf{Comparison of three computational methods.}}
\begin{tabular}{@{}lccc@{}}
\toprule
 & \textbf{TTM} & \textbf{TDBE} & \textbf{EMLMD} \\
\midrule
Electronic subsystem & Thermalized & Non-thermal & Thermalized \\
Phonon subsystem & Thermalized & Thermalized & Non-thermal \\
e--ph coupling resolution & Averaged & Phonon-resolved & Phonon-resolved \\
e--ph coupling evolution & Fixed & Fixed & Time-dependent \\
ph--ph coupling & Neglected & Neglected & Considered \\
\bottomrule
\end{tabular}
\end{table*}

\section{Discussion} \label{sec:discussion}

The TTM, TDBE, and EMLMD approaches represent three distinct modelling paradigms for excited-state dynamic simulations, each built upon unique physical assumptions and numerical approximations. While all three methods enable the description of coupled electron--phonon dynamic evolution, their valid regimes of applicability differ substantially due to distinct theoretical constraints and approximations, as summarized in Table~1.

Although modified TTM variants have been developed to partially account for non-thermal electron distributions in special cases \cite{Carpene2006, Uehlein2022}, the standard TTM framework is fundamentally designed for coupled dynamic evolution of thermalized electronic and phonon subsystems. This method prioritizes macroscopic statistical thermodynamic properties over atomic-scale fast dynamics. It is therefore well suited for dynamic evolution on extended timescales, but inherently limited in resolving sub-picosecond nonequilibrium and transient atomistic processes.

In contrast, the TDBE method specializes in capturing non-thermal electronic dynamics. In addition to its intrinsic capability to incorporate phonon--phonon interactions and non-thermal phonon effects, TDBE exhibits unique superiority in scenarios where individual electronic behavior dominates, including carrier transport, photoionization, and exciton formation, where electrons cannot be approximated as a thermalized ensemble. The method is particularly well-suited for low-excitation or ultrashort-timescale processes, where lattice displacement and structural perturbation remain marginal.

Distinctively, the EMLMD method is specifically established to tackle strong photoexcitation scenarios characterized by substantial atomic motion and prominent nonadiabatic dynamics, exemplified by photoinduced phase transitions. Conventional theoretical simulations have long suffered from inherent spatiotemporal mismatches with experimental observations of light-induced structural phase transitions. Experimentally resolved structural evolution typically proceeds on the picosecond-to-nanosecond timescale, whereas standard rt-TDDFT simulations are restricted to hundreds of femtoseconds. Moreover, nucleation and growth processes during the phase transition of realistic materials are spatially heterogeneous and dispersed, which cannot be adequately captured by conventional small cell calculations. The EMLMD framework effectively resolves these critical bottlenecks while maintaining first-principles level accuracy.

Although the current implementation is limited to systems with quasi-thermalized electron distributions, its physical generality and applicability can be further extended. Incorporation of sophisticated electron distribution functions fitted to rt-TDDFT trajectories can significantly broaden the accessible electronic phase space in excited states. Furthermore, integration of machine learning neural networks enables intelligent training and automated evolution of electronic parameters, achieving accurate excited-state dynamical propagation and faithful reproduction of rt-TDDFT nonadiabatic effects. Scaling of the EMLMD toward larger-scale systems will further allow the exploration of complex physical processes, including defect-modulated dynamics and heterogeneous phase transitions.

\section{Conclusion} \label{sec:conclusion}

In this work, we develop a machine learning-based excited-state molecular dynamics framework that incorporates electronic order parameters (such as $T_e$) as independent dimensions to accurately reconstruct excited-state PES. The time evolution of electronic characteristics is rigorously derived from high-precision rt-TDDFT benchmark trajectories, which are obtained from small-scale first-principles calculations. This approach intrinsically captures the excited-state electron--phonon coupling matrix elements and inherently incorporates phonon--phonon anharmonic interactions, enabling the construction of excited-state PES with first-principles accuracy. We comprehensively validate the reliability and universal applicability of EMLMD by two representative photoinduced structural transformation processes: the photoinduced crystalline phase transition of bismuth and photoamorphization of selenium. For bismuth structural transformation, our large-scale simulations provide time- and $\mathbf{q}$-resolved phonon occupation information and directly visualize the competitive interplay between coherent photoexcited phonons and disordered thermal phonons. For selenium photoamorphization, the framework simultaneously resolves atomistic microscopic dynamics and global structural transformation characteristics. Systematic comparison with the conventional TTM and TDBE methods further highlights the unique advantages and broad prospects of the newly developed EMLMD framework for probing the underlying dynamic processes and microscopic mechanisms of light-induced phenomena in functional materials.

\par\addvspace{4.0ex}
{\centering\normalfont\bfseries\MakeUppercase{Methods}\par}
\vspace{2.0ex}

\newcommand{\methodsubsection}[1]{%
  \par\addvspace{2.6ex}%
  \noindent\parbox{\columnwidth}{\centering\normalfont\bfseries #1\par}%
  \par\nobreak\vspace{1.6ex}%
}

\methodsubsection{Network training and dynamics based on the MLIP}

The generation of training data is based on a series of AIMD trajectories, generated by the SIESTA \cite{Soler2002} software package. The initial structures of the trajectories include the A7 phase, the SC phase, as well as some perturbed structures along the phase transition pathway. The trajectory dynamics cover electron temperatures ranging from $300\text{ K}$ to $24{,}000\text{ K}$, while the ionic temperatures span a subset from $100\text{ K}$ to $500\text{ K}$. Over $18{,}000$ data points were extracted from all AIMD trajectories for network training. The network architecture was modified from the open-source GPTFF \cite{Xie2024} software package by altering certain network modules to enable the recognition of more detailed structural changes. Additionally, the parameter dimensions for electronic information were expanded to allow the inclusion of labels such as, but not limited to, electron temperature. The dynamics simulation was performed using the Python ASE package under the NVT ensemble. The ionic temperature was maintained at $100\text{ K}$, with a timestep of $0.2\text{ fs}$ for a total of $20{,}000$ steps.

\methodsubsection{First-principles excited-state molecular dynamics simulations}

First-principles non-adiabatic molecular dynamics simulations were performed using time dependent ab initio package (TDAP) \cite{Lian2018, You2020}, where the electron dynamics are captured within the framework of real-time time dependent density functional theory (rt-TDDFT) and the coupled motion of electrons and nuclei is described by Ehrenfest dynamics \cite{Curchod2013}. The bismuth crystal is simulated with $2 \times 2 \times 2$ supercell, containing 32 atoms. The initial lattice dimensions of the supercell are $13.36\text{ \AA}$, $9.20\text{ \AA}$ and $9.63\text{ \AA}$. A numerical atomic-orbital basis set with double-zeta polarization (DZP) and Troullier--Martins pseudopotentials is adopted. We employ the PBE version generalized gradient approximation (GGA) \cite{Perdew1996} to describe the electronic exchange correlation interaction of the total energy. An auxiliary real-space grid equivalent to a plane-wave cutoff of $150\text{ Ry}$ is used and the Brillouin zone is sampled by a $2 \times 3 \times 3$ $\mathbf{k}$-mesh. A linearly polarized laser pulse is applied with a Gaussian-enveloped waveform $E(t)=E_0\cos(\omega t)\exp[-(t-t_0)^2/(2\sigma^2)]$, where the photon energy $\hbar\omega$ is $1.55\text{ eV}$, the field intensity $E_0=0.386\text{ V/\AA}$ and the time to the peak $t_0=32\text{ fs}$. The time step of $0.04\text{ fs}$ is applied for both electron and ion motions, evolving in a constant pressure-energy (NPE) ensemble, where pressure is controlled by Parrinello--Rahman method \cite{Parrinello1981}.

\methodsubsection{Band structure and phonon calculations}

Band structures are calculated via Vienna Ab Initio Simulation Package (VASP) \cite{Kresse1996} adopting the projector augmented wave (PAW) method. Phonon frequency and eigenvectors are computed via density-functional-perturbation theory (DFPT) with the PHONOPY \cite{Togo2015}.

\end{document}